\documentstyle[aps,amssymb]{revtex}
\begin{document}
\title{Global time asymmetry as a consequence of a wave packets theorem}
\author{Mario A. Castagnino, Jorge Gueron}
\address{Instituto de Astronom\'{\i}a y F\'{\i}sica del Espacio\\
Casilla de correos 67 sucursal 28\\
1428 Buenos Aires, Argentina.}
\author{Adolfo R. Ord\'{o}\~{n}ez}
\address{Instituto de F\'{\i}sica Rosario.\\
Av. Pellegrini 250,\\
2000 Rosario, Argentina.}
\date{November 26, 1999}
\maketitle

\begin{abstract}
When $t\rightarrow \infty $ any wave packet in the liouvillian
representation of the density matrices, becomes a Hardy class function from
below. This fact, in the global frame of Reichenbach diagram, is used to
explain the observed global time asymmetry of the universe.
\end{abstract}

\section{Introduction}

Many authors consider time asymmetry as having a global origin \cite{Global}%
. The Reichenbach branched diagram has being used to explain this global
phenomenon \cite{Reichenbach}. Here we make a resume of this idea.
Reichenbach diagram is the combination of all the scattering processes
within the universe evolution, beginning at the initial global instability
of the universe, which is considered as the source of all energy. In each
scattering process, the energy of the incoming states is used to produce
unstable states that decay, originating the outgoing ones. So, the outgoing
lines of the diagram can always be considered as evolutions from an unstable
state towards equilibrium. Essentially, the whole diagram can be considered
as having only outgoing lines. In fact, the incoming lines in each
scattering are outgoing lines of a previous process. Moreover, the incoming
lines of a scattering system ``S'' cannot be considered as {\it spontaneous
evolutions}, since they just show the pumping of energy from a precedent
process, that is really coupled with the scattering system ``S'', making
these incoming lines representatives of{\it \ non spontaneous} or forced
evolutions. On the contrary, spontaneity clearly characterizes outgoing
lines. Considered as a whole, Reichenbach diagram symbolize the asymmetrical
flow of all the energy within the universe from its initial instability
towards a final equilibrium state, resolving this flow as a sum of
scattering processes. Then, time asymmetry (which cannot be explained as a
consequence of the local time-symmetric physical{\it \ laws}), can be easily
explained as a consequence of the time asymmetry of the{\it \ object-universe%
}.

On the other hand, time asymmetry can be considered as the consequence of
the existence of a time-asymmetric space of physical admissible states \cite
{CL}. For causality reasons, this space was associated with the space of
Hardy class functions from below \cite{ABK}. If all the lines in the
Reichenbach diagram would be electromagnetic waves this choice would be
natural, since the outgoing waves of electromagnetic scattering can be
represented as states belonging to the just mentioned Hardy space \cite{LP}.
This conclusion can be extended to all hyperbolic scatterings but not to the
parabolic equation of non relativistic quantum mechanics. Nevertheless, we
will show that far from the scatterer, the outgoing lines belong to a Hardy
space if the physical admissible states are wave packets. Then, all the
spontaneously evolving states of the universe would belong to such a type of
space, that we will call $\Psi _{-}.$ The unphysical non spontaneous
time-inverted states would belong to a space $\Psi _{+}=K\Psi _{-}\neq \Psi
_{-},$ where $K$ is Wigner time-reversal operator \cite{Messiah} (\footnote{%
In the usual popularization language $\Psi _{-}$ would be the space of the
spontaneous evolutions: the sugar lump solving in the coffee, or the
elephant breaking the crystal shop. While $\Psi _{+}$ will be the space of
impossible (or better non-spontaneous) evolutions: the sugar lump
concentrating in the coffee, or the elephant reconstructing the crystal shop.%
}). The arrow of time would be the consequence of this asymmetry.

\section{The Baker's transformation}

As a dydactical introduction to the subject we will show how spaces $\Psi
_{-}$ and $\Psi _{+}$ appears in the famous Baker transformation.

Let us consider the unit square $S=\left[ 0,1\right] \times \left[
0,1\right] $ with its restricted Lebesgue measure $\mu $ and the Baker's
transformation: 
\begin{equation}
B(x,y)=\left\{ 
\begin{array}{c}
(2x,\frac{1}{2}y)\text{ if }0\leq x\leq \frac{1}{2} \\ 
(2x-1,\frac{1}{2}y+\frac{1}{2})\text{ if }\frac{1}{2}\leq x\leq 1
\end{array}
\right.  \label{2.1}
\end{equation}

Let us consider the {\it independent and generating partition} of $B$ \cite
{Courbage}, i. e., the partition of the unit square into its left and right
halves or ``vertical'' rectangules, $P=\left\{ \Delta _{1},\Delta
_{2}\right\} $. ``Independence'' of the partition $P$ with respect to $B$
means that 
\begin{equation}
\mu \left( \bigcap\limits_{n=-m_{1}}^{m_{2}}B^{n}\left( \Delta
_{i_{n}}\right) \right) =\prod\limits_{n=-m_{1}}^{m_{2}}\mu \left( \Delta
_{i_{n}}\right)  \label{2.3}
\end{equation}
where $\Delta _{i_{n}}\in P$. The ``generating'' character of the partition $%
P$ with respect to $B$ means that any Borel measurable set of the unit
square can be obtained by forming {\it countable} unions and intersections
of sets of the form 
\[
\bigcap\limits_{n=-m_{1}}^{m_{2}}B^{n}\left( \Delta _{i_{n}}\right) \;\text{%
with }\Delta _{i_{n}}\in P 
\]

Let $U:L^{2}(S,\mu )\rightarrow L^{2}(S,\mu )$ be the unitary map 
\begin{equation}
\left( U^{n}\rho \right) (w)=\rho \left( B^{-n}(w)\right)  \label{2.4}
\end{equation}
where $\rho \in L^{2}(S,\mu ),$ $w\in S$. As is well known, $U$ has a
countable uniform Lebesgue spectrum, and therefore \cite{Cotlar} there is a 
{\it system of imprimitivity} in $L^{2}(S,\mu )$ based on ${\Bbb Z}$ for the
group $\left\{ U^{n}:n\in {\Bbb Z}\right\} .$ In other words, there is a
spectral measure $E$ defined on ${\Bbb Z}$ and taking its values in the set
of the orthogonal projection operators of $L^{2}(S,\mu ),$ such that 
\begin{equation}
\forall n,m\in {\Bbb Z}:U^{-n}E_{m}U=E_{m+n}  \label{2.5}
\end{equation}
where $E_{m}=E\left( \left\{ m\right\} \right) $, and $E_{m+n}=E\left(
\left\{ m+n\right\} \right) $. Then, we can define the {\it Age operator }%
\cite{Courbage} by 
\begin{equation}
A=\int\limits_{{\Bbb Z}}ndE=\sum\limits_{n=-\infty }^{+\infty }nE_{n}
\label{2.6}
\end{equation}

As a consequence of eqs. (\ref{2.5}) and (\ref{2.6}), and taking into
account the properties of $E$, we have 
\begin{equation}
\forall n\in {\Bbb Z}:U^{-n}AU=A+nI  \label{2.7}
\end{equation}

The age operator $A$ has ${\Bbb Z}$ as uniform spectrum with countable
multiplicity. In fact, the functions 
\[
\alpha _{0}(w)=%
%TCIMACRO{
%\QATOPD\{ . {+1\text{ if }w\in \Delta _{1}}{-1\text{ if }w\in \Delta _{2}} }
%BeginExpansion
{+1\text{ if }w\in \Delta _{1} \atopwithdelims\{. -1\text{ if }w\in \Delta _{2}}%
%EndExpansion
\]
together with its transformed by $U^{n}:\alpha _{n}=U^{n}(\alpha _{0}),$ and
all their finite products 
\[
\alpha _{F}=\alpha _{n_{1}}...\alpha _{n_{r}} 
\]
($F=\left\{ n_{1},...,n_{r}\right\} $ $n_{j}\in {\Bbb Z}$) constitute an
orthonormal eigenbasis $\left\{ \alpha _{F}:F\subset {\Bbb Z}\right\} $ of $%
A, $ being each $\alpha _{F}$ a ``state'' of age $n=\max \left\{
n_{1},...,n_{r}\right\} $, in the sense that 
\begin{equation}
A\alpha _{F}=n\alpha _{F}  \label{2.8}
\end{equation}
(Clearly, there are countably many states corresponding to each eigenvalue
or age $n,$ showing its multiplicity). Because of eq. (\ref{2.5}), $U$ act
as a {\it right bilateral shift} \cite{Cotlar} on this basis:\ 
\begin{equation}
U\alpha _{F}=\alpha _{F+1}  \label{2.9}
\end{equation}
(where $F+1:=\left\{ n_{1}+1,...,n_{r}+1\right\} $). The eqs. (\ref{2.7})
and (\ref{2.9}) show how the age brought about by the dynamical evolution
and growing with it matches with the progress of external (or observer's)
time labelling the dynamical group.

Now, we can decompose $L^{2}(S,\mu )$ as a direct sum of two Hardy functions
spaces \cite{Cotlar} $H_{+}^{2}$ and $H_{-}^{2}$ such that: 
\begin{equation}
H_{+}^{2}=\left\{ \rho :\rho \in L^{2}(S,\mu )\wedge \rho
=\sum\limits_{F\subset {\Bbb Z}^{+}}a_{F}\alpha _{F}\right\}  \label{2.10}
\end{equation}
having only non null Fourier coefficients $a_{F}\in {\Bbb C}$ for positive
indices, and 
\begin{equation}
H_{-}^{2}=\left\{ \rho :\rho \in L^{2}(S,\mu )\wedge \rho
=\sum\limits_{F\subset {\Bbb Z}_{0}^{-}}a_{F}\alpha _{F}\right\}
\label{2.11}
\end{equation}
having only non vanishing Fourier coefficients for negative (or zero)
indices.

Then, it is obvious that: 
\begin{equation}
UH_{+}^{2}\subset H_{+}^{2},\text{ }UH_{-}^{2}\subset H_{+}^{2}\oplus
H_{-}^{2}=L^{2}(S,\mu )\neq H_{-}^{2}  \label{2.12}
\end{equation}
and that: 
\begin{equation}
\lim_{n\rightarrow +\infty }U^{n}(H_{+}^{2}\oplus H_{-}^{2})=H_{+}^{2}
\label{2.13}
\end{equation}

We will describe this fact by saying that the states belonging to $H_{+}^{2}$
are ``stable towards the future'' under the induced evolution $U$, while the
those belonging to $H_{-}^{2}$ are ``unstable'' (\footnote{%
Of course, with repect to the inverse evolution $U^{-1}$ ``towards the
past'' we must reverse these terms.}). In this way, when acted on by $U^{n},$
any function belonging to $L^{2}(S,\mu )$ will end in space $H_{+}^{2}$ in
the ``far future'' (precisely when $n\rightarrow +\infty ).$

In the next section we will consider the ``quantum version'' of what we have
said above.

\section{Pure states and the hamiltonian}

Let us begin considering just pure states $|\psi \rangle ,$ belonging to a
Hilbert space ${\cal H}$, of a quantum system with hamiltonian $H$, such
that: 
\begin{equation}
H|\omega ,n\rangle =\omega |\omega ,n\rangle  \label{4.1}
\end{equation}
where $0\leq \omega <\infty $ or $\omega \in {\Bbb R}^{+}$and $n$ belongs to
a set of indices $N$, which is the same for any $\omega $ (for didactical
reasons we will assume that this set is numerable, and therefore the index $%
n $ will be discrete). Thus, $H$ can be considered as a typical scattering
hamiltonian just endowed with an absolutely continuous and uniform energy
spectrum. Precisely, there is a nuclear space $\Phi $ and a rigging of it
with ${\cal H}$ 
\[
\Phi \subset {\cal H}\subset \Phi ^{\times } 
\]
such that 
\[
\{|\omega ,n\rangle :\omega \in {\Bbb R}^{+}\wedge \;n\in N\}\subset \Phi
^{\times } 
\]
(we wil denote by $\Phi ^{\times }$ the antidual space of $\Phi ,$ composed
of all continuous antilinear functionals on $\Phi $, and by $\Phi ^{\prime }$
its dual) is a generalized eigenbasis of $H$ \cite{Gelfand} \cite{Manolo} in
the sense that: 
\begin{equation}
\forall \varphi ,\psi \in \Phi :\left\langle \varphi |\psi \right\rangle
=\sum_{n}\int_{0}^{\infty }d\omega \langle \varphi |\omega ,n\rangle \langle
\omega ,n|\psi \rangle  \label{4.1.1}
\end{equation}
where the l. h. s. means the scalar product in ${\cal H}$ (antilinear in its
left factor), while in the r. h. s. $\langle \varphi |\omega ,n\rangle $
means the evaluation of the antilinear functional $|\omega ,n\rangle $ on $%
\varphi $, and $\langle \omega ,n|\psi \rangle $ is the evaluation of the
linear functional $\langle \omega ,n|$ on $\psi $. This justifies Dirac's
notation: 
\begin{equation}
|\psi \rangle =\sum_{n}\int_{0}^{\infty }d\omega |\omega ,n\rangle \langle
\omega ,n|\psi \rangle  \label{4.2}
\end{equation}

Moreover, let us consider that real physical states are wave packets,
mathematically modelled by Schwarz functions of $\omega \in {\Bbb R}^{+},$
for each value of $n,$ so: 
\begin{equation}
f(\omega )=\langle \omega ,n|\psi \rangle \in {\cal S}^{+}={\cal S}({\Bbb R}%
^{+})  \label{4.3}
\end{equation}
(${\cal S}^{+}$ is the space of all infinite differentiable complex-valued
functions defined on $\left[ 0,+\infty \right) $, such that converge to zero
for $\omega \rightarrow +\infty $ faster than the inverse of any polynomial).

Taking into account all the values of $n$ we can say that: 
\begin{equation}
f(\omega )=\langle \omega ,n|\psi \rangle \in \bigoplus_{n}{\cal S}_{n}^{+}
\label{4.4}
\end{equation}

This mathematical model is adopted for the following reasons:

1.- It is clear that we do not find infinite energies in nature, and so $%
\langle \omega ,n|\psi \rangle $ must somehow go to zero when $\omega
\rightarrow +\infty .$

2.- In order to use derivatives in our calculations it is not enough to
postulate that the states belong to a Hilbert space. They must be
representable by differentiable functions. We postulate that they are
infinitely differentiable. After all, we cannot find an experimental
contradiction to this assumption.

3.- But since these functions must also be square integrable, we can take
for granted that they go to zero when $\omega \rightarrow +\infty .$ We
postulate that they go to zero faster that the inverse of any polynomial.

Of course, we are free to choose other spaces instead of ${\cal S}^{+},$ but
it is evident that ${\cal S}^{+}$ is the simplest model endowed with all the
usual properties of wave packets (that's why the same choice is made in \cite
{Bogo}).

\section{Mixed states and the liouvillian}

We will use the notation of ref. \cite{CGG}. Then, the Liouville operator
reads. 
\begin{equation}
{\Bbb L}=[H,.]=H\times I-I\times H  \label{5.1}
\end{equation}

Let us consider the space of ``density matrices'' ${\cal L=H}\otimes {\cal H}
$, the rigged Hilbert space 
\[
\Phi \otimes \Phi \subset {\cal H}\otimes {\cal H}\subset \Phi ^{\times
}\otimes \Phi ^{\prime } 
\]
and the generalized basis: 
\begin{equation}
\left\{ |\omega ,n\rangle \langle \omega ^{\prime },n^{\prime }|:\omega
,\omega ^{\prime }\in {\Bbb R}^{+}\wedge n,n^{\prime }\in N\right\}
\label{5.1'}
\end{equation}
(where $|\omega ,n\rangle \langle \omega ^{\prime },n^{\prime }|=|\omega
,n\rangle \otimes \langle \omega ^{\prime },n^{\prime }|\in \Phi ^{\times
}\otimes \Phi ^{\prime }$).

Let us define the Riezs indices \cite{CGG}: 
\[
\nu =\omega -\omega ^{\prime },\text{ }-\infty <\nu <\infty 
\]
\begin{equation}
\sigma =\frac{1}{2}(\omega +\omega ^{\prime }),\;\text{ }\frac{|\nu |}{2}%
\leq \sigma <\infty  \label{5.2}
\end{equation}

It will be convenient to label the basis (\ref{5.1'}) as: 
\begin{equation}
|\omega ,n\rangle \langle \omega ^{\prime },n^{\prime }|=|\nu ,\sigma
,n,n^{\prime })  \label{5.3}
\end{equation}

Then: 
\begin{equation}
{\Bbb L}|\nu ,\sigma ,n,n^{\prime })=\nu |\nu ,\sigma ,n,n^{\prime })
\label{5.4}
\end{equation}

So 
\begin{equation}
\left\{ |\nu ,\sigma ,n,n^{\prime }):\nu \in {\Bbb R}\wedge \frac{|\nu |}{2}%
\leq \sigma <\infty \wedge n,n^{\prime }\in N\right\}  \label{5.4.1}
\end{equation}
is a generalized eigenbasis of the liouvillian, being $\nu $ the
corresponding generalized eigenvalue and $\sigma ,$ $n,$ $n^{\prime },$ the
``degeneration indices''. From (\ref{5.2}) we see that $\nu \in {\Bbb R}$,
while $n$ and $n^{\prime }\in {\Bbb N}$ , and $\sigma \in {\Bbb R}^{+}$, and
these spaces have the same cardinality for any $\nu $. So ${\Bbb L}$ has
uniform Lebesgue spectrum ${\Bbb R}$.

In the basis (\ref{5.4.1}) ``the $\nu $-wave function'' reads: 
\begin{equation}
\rho (\nu )=(\rho |\nu ,\sigma ,n,n^{\prime })  \label{5.5}
\end{equation}

Following the ideas of the previous section (\footnote{%
And taking into account that we are considering mixed states as ``density
matrices'' identified with {\it tensor products} of pure states.}), it is
physically justified to suppose that these functions are sums of products of
functions $f(\omega )\in {\cal S}\left( {\Bbb R}^{+}\right) $, namely: 
\begin{equation}
f(\omega )g(\omega ^{\prime })=f\left( \sigma +\frac{\nu }{2}\right) g\left(
\sigma -\frac{\nu }{2}\right)  \label{5.6}
\end{equation}

They will have infinite derivatives with respect to $\nu ,$ since $f$ and $g$
are infinitely differentiable. Moreover, $\rho (\nu )$ goes to zero when $%
\nu \rightarrow \pm \infty $ faster than the inverse of any polynomial since
this is a property of $f$ and $g.$ Thus, for any $\sigma ,$ $n,$ $n^{\prime
} $ : 
\begin{equation}
\rho (\nu )=(\rho |\nu ,\sigma ,n,n^{\prime })\in {\cal S}\left( {\Bbb R}%
\right)  \label{5.7}
\end{equation}

\section{The Age operator}

Since ${\Bbb L}$ has uniform Lebesgue spectrum ${\Bbb R}$, there is a system
of imprimitivity in ${\cal L}$ based on ${\Bbb R}$ for the group $\left\{
U_{t}:t\in {\Bbb R}\right\} .$ In other words, there is a spectral measure $E
$ defined on ${\Bbb R}$ and taking its values in the set of the orthogonal
projection operators of ${\cal L}$ \cite{Cotlar}$,$ such that 
\begin{equation}
\forall t,s\in {\Bbb R}:U_{t}^{-1}E_{s}U_{t}=E_{t+s}  \label{6.1.1}
\end{equation}
where $E_{s}=E\left( \left( -\infty ,s\right] \right) $, and $%
E_{t+s}=E\left( \left( -\infty ,t+s\right] \right) $. Then we can define the 
{\it Age operator }\cite{Misra} by 
\begin{equation}
A=\int\limits_{{\Bbb R}}t\,dE  \label{6.1.2}
\end{equation}

As a consequence of eqs. (\ref{6.1.1}) and (\ref{6.1.2}), and taking into
account the properties of $E$, we have 
\begin{equation}
\forall t\in {\Bbb R}:U_{t}^{-1}AU_{t}=A+tI
\end{equation}

The age operator $A$ has ${\Bbb R}$ as a uniform Lebesgue spectrum. In fact,
for the physical states we have:

\begin{equation}
{\Bbb A}\rho (\nu ,\sigma ,n,n^{\prime })\doteq i\frac{\partial }{\partial
\nu }\rho (\nu ,\sigma ,n,n^{\prime })|_{\sigma ,n,n^{\prime }=const.}
\label{6.1}
\end{equation}
that is equivalent to the commutation relation 
\begin{equation}
\lbrack {\Bbb A},{\Bbb L}]=i  \label{6.2}
\end{equation}
(${\Bbb A}$ and ${\Bbb L}$ have essentially the same commutation relation as
position and momentum operators $q$ and $p$)$.$ Then $\widehat{\rho }%
(a,\sigma ,n,n^{\prime }),$ the Fourier transform in variables $\nu
\leftrightarrow a$ of $\rho (\nu ,\sigma ,n,n^{\prime }),$ is an eigenvector
of ${\Bbb A}$, precisely: 
\begin{equation}
{\Bbb A}\widehat{\rho }(a,\sigma ,n,n^{\prime })=a\widehat{\rho (}a,\sigma
,n,n^{\prime })  \label{6.2.1}
\end{equation}

Moreover $\widehat{\rho (}a,\sigma ,n,n^{\prime })\in {\cal S}\left( {\Bbb R}%
\right) $ in the variable $a$ since it is the Fourier transform of $\rho
(\nu ,\sigma ,n,n^{\prime }).$ Then, the time evolution of $\widehat{\rho (}%
a,\sigma ,n,n^{\prime })$ reads: 
\begin{eqnarray}
e^{-i{\Bbb L}t}\widehat{\rho (}a,\sigma ,n,n^{\prime }) &=&e^{t\frac{%
\partial }{\partial a}}\widehat{\rho (}a,\sigma ,n,n^{\prime })=  \nonumber
\\
&=&\widehat{\rho (}a,\sigma ,n,n^{\prime })+t\frac{\partial }{\partial a}%
\widehat{\rho (}a,\sigma ,n,n^{\prime })+\frac{t^{2}}{2!}\frac{\partial ^{2}%
}{\partial a^{2}}\widehat{\rho (}a,\sigma ,n,n^{\prime })+...=  \nonumber \\
&=&\widehat{\rho (}a+t,\sigma ,n,n^{\prime })  \label{6.3}
\end{eqnarray}

Thus, ${\Bbb L}$ is the generator of the time translations, and $\widehat{%
\rho (}a,\sigma ,n,n^{\prime })$ increase its age as $a\rightarrow a+t$
becoming $\widehat{\rho (}a+t,\sigma ,n,n^{\prime }).$ This fact justify the
name given to ${\Bbb A}$. But eq. (\ref{6.3}) tell us that during its time
evolution, the wave packet $\widehat{\rho (}a,\sigma ,n,n^{\prime })\in 
{\cal S}\left( {\cal {\Bbb R}}\right) $ do not change its shape, being
merely shifted to the left. So, in the basis (\ref{5.4.1}) all physical
states are wave packets at any time, and moreover verify: 
\begin{equation}
\lim_{t\rightarrow +\infty }\widehat{\rho (}a+t,\sigma ,n,n^{\prime })=0
\label{6.4}
\end{equation}
because functions in Schwarz space go to zero when its variable goes towards
the infinite .

\section{The theorem}

The quantum version of the Baker's transformation example would be as
follows. Let us consider a quantum system whose states $\rho $ are ``density
matrices'' belonging to a Hilbert-Liouville space ${\cal L}$. Let ${\Bbb L} $
be the Liouville superoperator in ${\cal L}$, assumed as having ${\Bbb R} $
as uniform Lebesgue spectrum. This amounts to say that the evolution
superoperator $\exp \left[ -i{\Bbb L}t\right] $ is a bilateral shift,
closely related with Hardy classes \cite{Rosenblum} \cite{Cotlar}.

We can decompose the space of physical states $\Psi =\Phi \otimes \Phi $ as $%
\Psi _{+}\oplus \Psi _{-},$ where 
\begin{equation}
\Psi _{+}=\left\{ \rho :\left[ \forall a<0:\widehat{\rho (}a,\sigma
,n,n^{\prime })=0\right] \right\}  \label{7.0}
\end{equation}
\[
\Psi _{-}=\left\{ \rho :\left[ \forall a>0:\widehat{\rho (}a,\sigma
,n,n^{\prime })=0\right] \right\} 
\]
are spaces of wave packets that are also Hardy class functions in the $\nu $%
-variable$,$ from $\{_{\text{below}}^{\text{above}}\}$. Then our theorem
states that:

{\bf Theorem}: {\it The limit of any physical state, when }$t\rightarrow
+\infty $, {\it belongs to }$\Psi _{-}.$

{\bf Proof: }We can decompose any $\widehat{\rho }(a)$ (abbreviation for $%
\widehat{\rho (}a,\sigma ,n,n^{\prime })$) as: 
\begin{equation}
\widehat{\rho }(a)=\widehat{\rho }_{+}(a)+\widehat{\rho }_{-}(a)  \label{7.1}
\end{equation}
such that: 
\[
\widehat{\rho }_{+}(a)=\widehat{\rho }(a)\text{ for }a>0,\text{ }\widehat{%
\rho }_{+}(a)=0\text{ for }a<0 
\]
\begin{equation}
\widehat{\rho }_{-}(a)=\widehat{\rho }(a)\text{ for }a<0,\text{ }\widehat{%
\rho }_{-}(a)=0\text{ for }a>0  \label{7.2}
\end{equation}

From eq. (\ref{6.4}) we see that when $t\rightarrow +\infty $, then $%
\widehat{\rho }_{-}(a+t)\rightarrow 0,$ and therefore $\widehat{\rho }%
(a+t)\rightarrow \widehat{\rho }_{+}(a+t)$. So, when $t\rightarrow +\infty $
these last two functions are zero for $a<0,$ and therefore, by the
Paley-Wiener theorem (\cite{Manolo}, page 47) the Fourier transform of $%
\widehat{\rho }(a),$ namely $\rho (\nu )$ belongs to $H_{-}^{2}.$ $\square $

So, we have proved

\begin{equation}
\lim_{t\rightarrow +\infty }e^{-i{\Bbb L}t}\Psi {\cal =}\Psi _{-}
\label{7.3}
\end{equation}
the analog of eq. (\ref{2.13}) for the wave packets, as announced.

\section{Conclusion}

As the typical distance among the scatterers is much bigger than the
characteristic dimension of the scatterers itself, most of the states can be
considered far from these scatterers. Therefore, most of the physical states 
{\it do} belong to space $\Psi _{-},$ thus explaining time-asymmetry (see
the introduction). Moreover, using the space of physical admissible states $%
\Psi _{-},$ most of the irreversible phenomena of nature can be foresee,
obtaining the same results as those of other formalisms (as
``coarse-graining'', Lindblad, etc. \cite{Fin}).

\end{document}